\documentstyle[aps,preprint,epsf]{revtex}
\def\be {\begin{equation}}
\def\ee {\end{equation}}
\def\vb {{\bar v}}
\def\ub {{\bar u}}

\def\dth{{{\partial \over {\partial \theta}}}}
\def\dxi{{{\partial \over {\partial \xi}}}}
\def\dph{{{\partial \over {\partial \phi}}}}
\def\beq{\begin{eqnarray}}
\def\eeq{\end{eqnarray}}
\def\phiIV{{\Xi_{3-\alpha, -3 + 3\alpha}}}
\def\J {{\mu}}

\begin{document}

\thispagestyle{empty}    

\title
{\bf  Additional analytically exact solutions for three-anyons
\footnote{Supported in part
by the U.S. Department of Energy
(D.O.E.) under cooperative agreement \# DE-FC02-94ER40818,
by Yonam foundation and by Korea Science and Engineering Foundation.}
}

\author{Chaiho Rim \thanks{Permanent address:
Chonbuk National University, Department of Physics, Chonju,
560-756, Korea.} }
\address{Center for Theoretical Physics,
Laboratory for Nuclear Science\\
and Department of Physics,
Massachusetts Institute of Technology \\
Cambridge, Massachusetts  02139 U.S.A.}

\maketitle

\thispagestyle{empty}    

\bigskip   
\begin{center}
MIT-CTP \#: 2494 ( hep-th/9512051 )\\
December  1995
\end{center}

\bigskip   

\begin{abstract}
We present new family of exact analytic solutions
for three anyons in a harmonic potential (or in  free space)
in terms of generalized harmonics on $S^3$,
which supplement the known solutions.
The new solutions satisfy the hard-core condition
when $\alpha={1\over 3},1$ ($\alpha$
being the statistical parameter)
but otherwise, have finite non-vanishing two-particle
colliding probability density,
which is consistent  with self-adjointness
of the Hamiltonian.
These  solutions, however,  do  not have
one-to-one mapping property
between bosonic and fermionic spectra.
\end{abstract}

\vfil

\newpage

Smooth interpolation
between bosonic and (spinless) fermionic spectra  has been a
very attractive idea and appeared  the concept of anyon \cite{myr1}
based on the homotopy group in two space dimensions.
The anyons are described effectively
by  Schr\"odinger equation of bosonic or  fermionic particles
with statistical gauge field of Aharonov-Bohm type.  One can also
derive \cite{csq} this equation of
motion from the second quantized
abelian Chern-Simons gauge theory.
It is believed that fractional quantum Hall effect
is an example of physical phenomenon of anyons \cite{wil1}.

On the other hand,
the detailed analysis of many anyons has not been successful.
Even for simple system such as harmonic oscillator (HO),
analytically known exact solutions
\cite{sol,chin,rim1,myr2}
fall short of the numerical result in \cite{num1}
which demonstarates smooth interpolation.
Only perturbation approach \cite{pert}
could reproduce the numerical result.
The main difficulty of the problem lies in the non-trivial
exchange property of anyons and hard-core
condition on its wavefunction.

In this Letter, we present  new family of
analytically exact solutions
of three particles in  HO or  in  free space.
These solutions supplement the presently known
analytically exact solutions but unfortunately
do not satisfy the hard-core condition
except at $\alpha=1/3,1$ ($\alpha$ is the statistical
parameter chosen as $0\le\alpha\le1$,
$\alpha$=0 at for boson and 1 for spinless fermion).
None the less, considering
the situation of absence of analytic solutions
for missing states, we think
these solutions will provide valuable information about anyons.

Our analysis is  done by adopting ordinary Hamiltonian
with anyonic particle-exchange property imposed on
wavefunction.
We use coordinates system following \cite{myr2}.
Coordinates of three particles are represented
as complex numbers, $z_a = x_a + i y_a$
where $a= 1,2,3$.  We separate out center of mass
(CM is irrelevant to the anyonic property) by using
CM  coordinate,
$Z = { (z_1 + z_2 + z_3)\over  \sqrt3}$ and
relative ones
$u={(z_1 + \eta\,  z_2 + \eta^2\, z_3)\over  \sqrt 3}$ and
$v={(z_1 + \eta^2\, z_2 + \eta\, z_3)\over  \sqrt 3}\,$
with $\eta = e^ {i 2\pi \over 3} $.

Relative motion (RM) of free particles is described
in terms of Hamiltonian (scaled as dimensionless),
$ H_{free} =
-{\partial \over \partial u}{\partial \over \partial  \bar u}
-{\partial \over \partial v}{\partial \over \partial  \bar v}
$.
We are considering systems with non-singular  potential
energy of the form, $V= V(r)$ where
$r= \sqrt{ u\bar u + v\bar v} \ge 0$. This includes the
free case, $V=0$ and
HO case, $V = r^2$.
This Hamiltonian is
invariant under exchange of any two particles. To see this
in our notation,
let us  denote the second and third particle
exchange operation, $(1,2,3) \to (1,3,2)$
as $E$, and cyclic operation, $(1,2,3) \to (2,3,1)$
as $P$.
Any two-particle exchange is represented in combination
of $E$'s  and $P$'s.
The definition of $u$ and $v$ shows that
$P: ( u,v)\to ( \eta^2 u, \eta v) $ and
$E:  (u, v) \to (v, u)$.
Therefore,  RM is trivially invariant under $P$ and $E$.

For separation  of variables,  we introduce
spherical coordinates,
$u= r \sin(\xi/2)\,\, e^{i ( \theta + {\phi  \over 2}) }$ and
$v = r \cos (\xi/2) \,\, e^{i (\theta -{\phi  \over 2}) }$.
$\xi$, $\theta$ and $\phi$ are angle variables
whose fundamental
domain is given as
$0 \le r$, $ 0\le \xi <{\pi \over 2}$,
$-{\pi \over 3}\le  \phi<{\pi \over 3}$, $0 \le \theta <2\pi$
since under $P$ and $E$, we have
$P :(r, \xi, \phi, \theta)
\to (r,\xi, \phi+{2\pi\over3}, \theta + \pi)$
and
$E: (r,\xi, \phi, \theta)
\to (r, \pi -\xi,  -\phi, \theta)$
respectively.
The angular domain is not $S^3$ but
${S^3 \over  Z_2 \times Z_3} $.
(Euler angles  on $S^3$ or on $SU(2)$
are defined
typically as $(\xi, \chi=-\phi,\psi \equiv 2\theta -2\pi)$
whose ranges are given as
$ 0\le  \xi <\pi$,
$0\le \chi<2\pi $, $-2 \pi  \le \psi <2\pi$).

Anyon wavefunction is defined to have  a phase $e^{i \alpha \pi}$
when any of two particles are interchanged.
This requires the wavefunction have the phase under $P$ and $E$,
\beq
&&E:\quad \Psi (r, \pi - \xi,
-\phi, \theta) = e^{i \alpha \pi}
\Psi (r, \xi, \phi, \theta)
\nonumber \\
&&P:\quad \Psi(r, \xi, \phi + {2 \pi \over 3}, \theta + \pi)
= e^{i 2 \alpha \pi}
\Psi(r, \xi, \phi, \theta)\nonumber \\
&&R: \quad
\Psi(r,\xi, \phi, \theta+ 2\pi)
= e^{i 6 \alpha \pi} \Psi(r, \xi, \phi, \theta)
\label{phase}
\eeq
The operation $R$, $2\pi$ rotation of $\theta$
corresponds to
a composite operation $EPEP$.

The Hamiltonian is written in ($r, \xi, \phi, \theta$)
coordinates as
\be
H = - {1\over {4r^3}} {\partial\over \partial r}
(r^3 {\partial \over \partial r} )
+ {1\over {4r^2}} M + V(r^2)
\label{Hr}
\ee
where $M$ is a Laplacian on $S^3$,
\be
M = - {4\over \sin \xi} \dxi \sin \xi  \dxi
+ {1 \over \sin ^2(\xi /2)}({1 \over 2 i} \dth + {1 \over i} \dph)^2
+ {1 \over \cos ^2(\xi/2) }({1 \over 2i} \dth - {1 \over i} \dph)^2 \,.
\label{M}
\ee
$M$ commutes with the Hamiltonian $H$.
The  relative angular momentum, $L$,
\be
L = u {\partial \over \partial u} + v { \partial \over \partial v}
- \ub {\partial \over \partial \ub} - \vb {\partial \over \partial \vb}
={1\over i} \dth\,,
\label{L}
\ee
is another  useful mutually commuting operator.
A simultaneous  eigenstate of $H$, $M$ and $L$
is given in factorized form,
$  \Psi_{E, \J, l} (r, \xi, \phi, \theta)
 = R_{E, \J}(r) \,\, \Xi_{\J, l} (\xi, \phi, \theta)
$.
$\Xi_{\J, l}$ is a harmonic on $S^3$,
whose two-particle analogue is $e^{\pm i\alpha \theta_{12}}$
where $0 \le \theta_{12} \le \pi $ is the angle of the relative
coordinate. $E$, $\J (\J + 2)$ and $l$ are
eigenvalues of $H$, $M$ and $L$ respectively.

Since one can easily solve the radial solution
once the harmonics are found
(for example, it is written as a
Bessel function of the first kind for free particle and
as a Laguerre for HO{}),
we concentrate on finding the angular part.
Boundary conditions  for $\Xi$
are  obtained
\cite{myr2,rim2}
from
Eq.~(\ref{phase}), consistently  with self-adjointness of $L$ and $M$,
\beq
&&\Xi(\xi, \phi, \theta=2\pi) =
e^{i6\alpha \pi}\,\, \Xi(\xi, \phi, \theta=0)
\nonumber\\
&&\Xi(\xi, \phi={\pi \over 3}, \theta=\pi) =
e^{i2\alpha \pi}\,\, \Xi(\xi, \phi=-{\pi \over 3}, \theta=0)
\nonumber\\
&&{\partial \over \partial \phi}\,\, \ln \Xi (\xi, \phi={\pi \over 3},
\theta=\pi)= {\partial \over \partial \phi} \,\,
\ln \Xi (\xi, \phi={\pi \over 3}, \theta=0)
\nonumber\\
&&\Xi(\xi= {\pi \over 2}, \phi, \theta)=
e^{-i\alpha \pi}\,\, \Xi (\xi = {\pi \over 2}, -\phi, \theta)
\nonumber\\
&&{\partial \over \partial \xi}
\,\, \ln\Xi(\xi ={\pi\over 2} ,\phi, \theta) =
-{\partial \over \partial \xi} \,\,
\ln \Xi (\xi={\pi \over 2}, -\phi, \theta)\,,
\label{b-cond}
\eeq
where last two identities hold for $0 <\phi \le \pi /3$.
In addition, any current across the boundary needs to be finite,
which requests
\be
\Xi(\xi=0, \phi, \theta) \,,\quad
\lim_{\Delta \phi \to 0} \Delta \phi
{\partial \over \partial \xi} \Xi(\xi ={\pi \over 2}, \Delta \phi
 , \theta)\,,\quad
{\partial \over \partial \phi } \Xi(\xi={\pi\over 2},
 \phi =\pm {\pi \over 3}, \theta)
\label{c-cond}
\ee
finite.
We note that even though hard-core condition
(vanishing at
$\xi\!=\! \pi /2$ and $\phi\!=\!0$) is
consistent with this boundary conditions,
it is not the unique choice for $\alpha \ne $ odd integer
as we will see promptly.

The harmonics are not   single- or double-valued
but multi-valued since  boundary conditions
contain $e^{i\alpha \pi}$ instead of ordinary $\pm1$.
To find the harmonics, we employ
ladder operators \cite{vilenkin}
satisfying the $su(2)$ Lie-algebra,
$[K_3, K_{\pm}] = \pm K_\pm$
and  $[K_+, K_-] =  2K_3$ where
\beq
&&K_+ = e^{i2 \theta} \{ i \dxi
- {1 \over  \sin \xi} \dph - {\cot \xi\over 2}   \dth \}
\nonumber \\
&&K_- = e^{-i2\theta} \{ i  \dxi
+ {1 \over  \sin \xi} \dph +{\cot \xi\over 2}  \dth \}
\nonumber \\
&&K_3= {1 \over2i} \dth
 = {1 \over 2} L \,,
\eeq
where $K_-^+= K_+$.
(Integration measure is  $\,\sin 2\xi d \xi d \phi d\theta$).
$K_\pm$ and $K_3$  are invariant under $P$ and $R$ but
get extra $-$  sign under $E$.
Noting that
$M= 4\{{ K_+K_- + K_-K_+ \over 2}   +  K_3^2\} $,
$\,\,\mu /2 $ is Casimir number for harmonics
and is semi-positive definite.

Well-known analytic solutions
called type $I$ ($II$)
\cite{sol,rim1} are reproduced in our approach
if we apply
$K_-^2$ ($K_+^2$) on  states which are
annihilated by $K_+^2 $ ($K_- ^2$).
Of course,
they  satisfy the hard-core condition for $0 \le \alpha \le 1$
and $\J = $ integer $ +3 \alpha \ge 0$
for type $I$
($\J = $ integer $ -3 \alpha \ge 3$
for type $ II$).

Newly found  solutions cannot be constructed this way.
They do not have states which
are annihilated by $K_+^2$ or $K_-^2$.
Suggestion of this new type has been made in \cite{chorim}.
Similar solutions but approximate ones
were considered in  \cite{chin}.
These have
$\J = $ integer $\pm \alpha \ge 2 $
($+$ $(-)$ for type $III$ $(IV)$).
(Note that the  convention for the class of the solution
in \cite{chin}  is
different from ours: $II$, $III$ and $IV$ corresponding
to our $III$, $IV$ and $II$ respectively).

A solution
with ($\J= 3 -\alpha,  l= -3 + 3\alpha)$ (type $IV$),
resulting in the interpolation between the fermionic
ground state and a bosonic excited state of HO,
is given as (using notations of $z\equiv v/r$ and $w \equiv u/r)
\,\,\,$
$ \phiIV  =
(z^3-w^3)^\alpha P_{3-\alpha, -3+3\alpha}\,\,\,
$
with
\be
P_{3-\alpha, -3+3\alpha}=
{ (\bar z+ \bar w)^{3-2\alpha}  \over
		 (z ^2 + zw + w^2 )^{\alpha} }
+ \text{ 2 more cyclic permutations } \,,
\ee
where 2  more added terms are obtained from the first term
if $z$ ($w$) is replaced  by $\eta z$ ($\eta^2 w)$, and
by $\eta^2 z$ ($\eta w$) respectively. Note that
for  $|w| < |z|$,  $\eta^3 w=w$
but $\eta^3 z \ne z$ because of
multi-valuedness of the factors in $\phiIV$.
The solution is equally put as
$
\Xi_{3-\alpha, -3+3\alpha}
=(z-w)^\alpha(\bar z + \bar w)^{3-2\alpha}
+(z-\eta w)^\alpha (\bar z + \bar \eta \bar w)^{3-2\alpha}
+(z-\eta^2 w)^\alpha (\bar z + \bar \eta^2 \bar w)^{3-2\alpha}
$
and more symbolically as
$\Xi_{3-\alpha, -3+3\alpha}
=S\{(z-w)^\alpha(\bar z + \bar w)^{3-2\alpha}
\}$.
Its explicit form is  given as
$\Xi_{3,-3} = 3(\bar z^3 + \bar w^3)$
in the bosonic limit, and
$\Xi_{2,0} = 3 (z\bar z - w \bar w)$
in the fermionic limit.

One can check that
$\Xi_{3-\alpha, -3+3\alpha}$
 is an  eigenfunction of $M$
using
$K_+ \bar z= - i w$,
$K_+ \bar w=  i z$,
$K_- z= - i \bar w$ and
$K_- w=  i \bar z$.
$\phiIV$ satisfies the boundary conditions
in Eqs.~(\ref{b-cond}, \ref{c-cond}), yet
does  not vanish but is finite
when two of three particles coincide ($z\!=\!w\!=\!1$),
except for  $3\alpha=$ odd integer:
\beq
\phiIV (z\!=\!w\!=\!1)= -(1-\eta)^\alpha\,
e^{i 2\pi \alpha \over 3} (1+ e^{i3\pi \alpha})\,.
\label{coin}
\eeq
We have used identities such as
$\lim_{\delta \to 0} (1+ (1-\delta) \eta) = e^{i {\pi \over 3}}$,
$\lim_{\delta \to 0} (1+ (1-\delta) \eta^2)
= e^{i {5\pi \over 3}}$
and their complex conjugates.
(Coordinates ($z_1$, $z_2$, $z_3$) is not suited
to proving Eq.~(\ref{coin}) due to
lack of faithful representation of $(-1)^\alpha$).
Two particles in the presence of the third
can  collide each other, as seen in two-particle case
in the context of self-adjoint extension
\cite{manuel}.
The chance
for simultaneous collision of three particles
is, however,  strictly prohibited except the bosonic limit
because radial part vanishes.

Will the enumeration of these solutions (FIG.~1,
whose details will appear
elsewhere) lead to smooth interpolation?
A solution is missing, interpolating
between ($\mu$=2, $l$=$-$2) boson and ($\mu$=3, $l$=1) fermion,
but this only implies that $\mu$$\ne$2+$\alpha$.
A real potential problem against the smooth interpolation
comes from the possibility of
a {\it two-to-one} correspondence.
For example,
two bosonic states
with ($\mu$=5, $l$=$-$1) and ($\mu$=3, $l$=$-$1)
are mapped into  a fermionic state
with ($\J$=4, $l$=2):
\beq
\Xi_{5-\alpha, -1+3\alpha} &=&
 S\{ (z- w)^{2+\alpha}(\bar z +\bar w)^{3-2\alpha}\}\,,
\\
\Xi_{3+ \alpha, -1+3\alpha} &=&
 S\{(\bar z - \bar w)^{2 - \alpha} (z+w)^{1+2\alpha}\}
\,.
\label{4sol}
\eeq
$\Xi_{5-\alpha, -1+3\alpha}$ is an extra solution
which has no counter-part to the numerical solution in Ref\cite{num1}.
It satisfies the boundary conditions in Eqs.~(\ref{b-cond},
\ref{c-cond}) and
is obtained by applying $K_+^2$ on
$\Xi_{5-\alpha,-5+3\alpha}=S\{ (z-w)^\alpha
(\bar z + \bar w)^{5-2\alpha}\}$.
Since that hard-core condition is satisfied at
$\alpha$=1/3, the relaxed boundary condition
cannot be the cause of this trouble.
It might be possible that the extra solution is an
isolated one which cannot be interpolated smoothly to
bosonic and fermionic limit satisfying the hard-core condition.
Even so, it does not eliminate the fact that there are
extra states beyond smooth interpolation.

For HO, energies of our solutions are linear in $\alpha$,
since the energy
is given as $\J+2+2n $ ($n$ is a non-negative integer and
$2n$ is due to the radial part).
However, as we impose the hard-core condition,
the wavefunctions will adjust themselves
and energies will have a slight non-linear dependence on $\alpha$.
This non-linear dependency should be  manifest
at the near  side of the fermionic states
of new solutions since all the solutions
satisfy the hard-core condition at $\alpha=1/3$.
In fact, this property is seen in
numerical \cite{num1},  perturbative
\cite{pert} and approximate estimation \cite{chin},
even though $\mu$
at $\alpha =1/3$  does not agree completely.

One can check that
reflection symmetry in $\mu$; $\alpha$ by
$2-\alpha$ \cite{rim1} for type $I$ and $II$,
and $\alpha$ by
$1-\alpha$ \cite{sen} for $III$ and $IV$,
although the latter is not perfect, being subject to
hard-core boundary condition.
We also expect that the tendency of linear dependence of $\mu$ on
$\alpha$ for $N$ anyons,
$({N(N-1) \over 2})\alpha$,
$({N(N-1) \over 2} -2)\alpha$,
$\cdots$,
$-({N(N-1) \over 2}-2)\alpha$,
and
$-({N(N-1) \over 2})\alpha$
up to an integer,
will persist if we relax the hard-core boundary condition.
This trend is shown in the numerical analysis
for four anyons \cite{num2}.

In summary, we present
four types of analytically exact normalizable
wavefunctions for three anyons, two of which are newly found.
The solutions are given in terms of
generalized harmonics on $S^3$,  with boundary
conditions consistent with anyonic exchange property,
self-adjointness of the Hamiltonian and no-infinite
current across the boundary.
The hard-core condition turns out to be an additional constraint
on the harmonics, which is obeyed by new family of solutions
at $\alpha$=1/3, 1.
The analytic expressions  for the whole solutions
do not give a  one-to-one correspondence
mapping between the bosonic and fermionic spectra,
whose way out is not seen at present.

The author has a great benefit from discussions with
Professors R. Jackiw, M. Rocek, J. Verbaarschot and I. Zahed,
and Drs. G. Amelino-Camelia and  K. Cho.

\begin{figure}
\epsfverbosetrue
\epsfxsize=270pt
\hbox to \hsize{\hfil{\epsffile{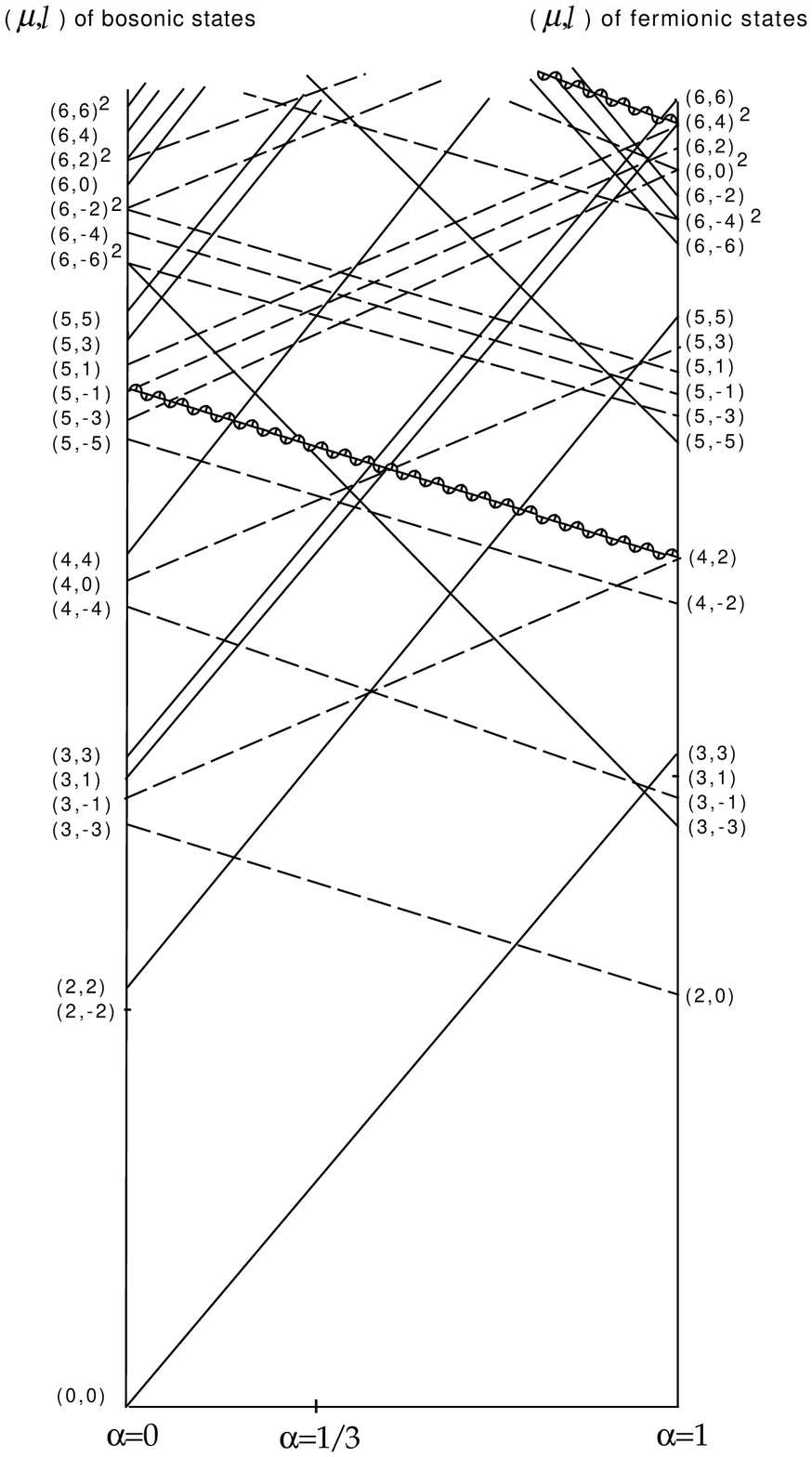}}\hfil}
\vspace*{0.33in}
\caption{ Four types of interpolation
between bosonic and fermionic states:
Schematic diagram for Casimir number ($\mu$) {\it v.s.}
statistical parameter ($\alpha$).
Solid lines represent Type $I$ and $II$, and dashed ones
$III$ and $IV$.
Wiggly lines are extras which prevent smooth
interpolation.
Exponents  on brackets denote degeneracy.}
\label{Fig. 1}
\end{figure}

\end{document}